\documentclass[epj]{webofc}
\usepackage[varg]{txfonts}   
\wocname{EPJ Web of Conferences}
\woctitle{ICNFP 2015}
\begin{document}
\selectlanguage{english}
\title{A new approach to analytic, non-perturbative, gauge-invariant QCD renormalization is described, with applications to high energy elastic pp-scattering.}
%
%

\author{ H. M. Fried\inst{1} \and
        P. H. Tsang\inst{1}\fnsep\thanks{\email{peter_tsang@brown.edu}} \and
        Y. Gabellini\inst{2} \and
        T. Grandou\inst{2} \and
        Y-M. Sheu\inst{1,2}
}

\institute{Brown University, Providence, RI 02912, USA
\and
           Institut Non Linéaire de Nice, Université Nice Sophia Antipolis, France          
}

\abstract{%
  A new non-perturbative, gauge-invariant model QCD renormalization is applied to high energy elastic pp-scattering. The differential cross-section deduced from this model displays a diffraction dip that resembles those of experiments. Comparison with ISR and LHC data is currently underway.}
\maketitle
\section{Introduction}
\label{intro}

We start with the Schwinger Generating Functional, GF, for QCD, with gluon operators in an Arbitrary, relativistic, gauge. The GF is rearranged in terms of a "Reciprocity Relation" under a "Gaussian Linkage Operation". The GF now depends upon two functionals of A,

\begin{equation}
\mathfrak{Z}_{QCD}[j,\bar{\eta},\eta]=\mathfrak{N}e^{-\frac{1}{2}\int \frac{\delta}{\delta A}\cdot D_c^{(0)}\cdot \frac{\delta}{\delta A}} \cdot e^{-\frac{i}{4}\int \bf{F}^2 + \frac{i}{2}\int A\cdot (-\partial^2)\cdot A}\cdot e^{i\int \bar{\eta}\cdot \bf{G}_c[A]\cdot \eta+\bf{L}[A]} |_{A=\int \bf{D}_c^{(0)}\cdot j}
\end{equation}

where the quark line, $\bf{G}_c[x,y|\bf{A}]= [m+\gamma\cdot (\delta-igA\tau)]^{-1}$
and virtual quark loop, $\bf{L}[\bf{A}]=\ln [1-i\gamma \bf{A}\tau_c[0]]$.

The GF can now can be rearranged into gauge-invariant form. This was overlooked for decades~\cite{FriedGabellini2010}.

Now combine with Fradkin expressions for the quark line, $\bf{G}_c[A]$, and quark loop, $\bf{L}[A]$. Efim S. Fradkin gave expressions for $\bf{G}_c[A]$ and $\bf{L}[A]$ in gaussian form~\cite{fradkin}. These are exact.

The $\bf{F}^2$ can be rewritten using Halpern's half century old expression~\cite{halpern}, 

\begin{equation}
e^{-\frac{i}{4}\int \bf{F}^2} = N \int d[\chi]e^{\frac{i}{4}\int \chi^2 + \frac{i}{2}\int F\cdot \chi}, 
\end{equation}

where $\chi_{\mu\nu}^a = -\chi_{\nu\mu}^a$.

With the $exp[- \frac{i}{4} \int \bf{F}^2]$ in the GF in Gaussian form, under $\chi$ fields, the relevant Gaussian Functional operations can be performed exactly. This corresponds to the summation of all Feynman graphs of gluons exchanged between quarks.

Then the explicit cancellation of all the gauge-dependent gluon propagators is obtained ~\cite{2}.

\section{Explicit Gauge Invariance}

A rearrangement can now be made to formally insure gauge-invariance, even though the GF still apparently contains gauge-dependent gluon propagators.

\begin{equation}
\mathfrak{Z}_{QCD}[j,\bar{\eta},\eta]=\mathfrak{N}e^{-\frac{1}{2}\int \frac{\delta}{\delta A}\cdot D_c^{(0)}\cdot \frac{\delta}{\delta A}} \cdot e^{-\frac{i}{4}\int \bf{F}^2 + \frac{i}{2}\int A\cdot (-\partial^2)\cdot A}\cdot e^{i\int \bar{\eta}\cdot \bf{G}_c[A]\cdot \eta+\bf{L}[A]} |_{A=\int \bf{D}_c^{(0)}\cdot j}
\end{equation}

Gives 2n-point functions: 

\begin{equation}
= \mathfrak{N}\int d[\chi]e^{\frac{i}{4}\int \chi^2  } e^{\mathfrak{D}_A^{(0)}}e^{\frac{i}{2}\int \chi \cdot \bf{F}+\frac{i}{2}\int A\cdot (D_c^{(0)})^{-1} \cdot A}G_c(1|gA)G_c(2|gA)e^{L[A]}|_{A=0}
\end{equation}

Then, 

\begin{equation}
\begin{split}
e^{\mathfrak{D}_a}F_1[A] &= exp[\frac{i}{2}\int \bar{Q}\cdot D_c^{(0)}\cdot (1-\bar{K}\cdot D_c^{(0)})^{-1}\cdot \bar{Q} - \frac{1}{2} tr \ln (1-D_c \cdot \tilde{K})] \\
&\quad \cdot\ exp[\frac{1}{2}\int A\cdot \bar{K}\cdot (1-D_c^{(0)}\cdot \bar{K})^{-1}\cdot A + i\int \bar{Q}\cdot (q-\bar{K}\cdot D_c^{(0)})^{-1}]
\end{split}
\end{equation}

where
\begin{equation}
D_c^{(0)}\cdot (1-\bar{K}\cdot D_c^{(0)})^{-1} = D_c^{(0)}\cdot [1-(\hat{K}+(D_c^{(0)})^{-1})\cdot D_c^{(0)}]^{-1}
=-(\tilde{K}_{\mu\nu}^{ab}+g f^{abc}\chi_{\mu\nu}^c)^{-1}= -\hat{\bf{K}}^{-1}
\end{equation}

\begin{equation}
\begin{split}
e^{\mathfrak{D}_A}F_1[A]F_2[A] &= exp[-\frac{i}{2}\int \bar{Q}\cdot \hat{\bf{K}}^{-1}\cdot \bar{Q}+ \frac{1}{2}tr \ln \hat{\bf{K}}+ \frac{1}{2} tr \ln(-D_c^{(0)})]  \\
&\quad \cdot\ exp[\frac{i}{2}\int \frac{\delta}{\delta A'}\cdot \bf{D}_c^{(0)}\cdot \frac{\delta}{\delta A'}]
\cdot exp[\frac{i}{2}\int \frac{\delta}{\delta A'}\cdot \hat{\bf{K}}^{-1}\cdot \frac{\delta}{\delta A'}- \int \bar{Q}\cdot\hat{\bf{K}}^{-1}\cdot \frac{\delta}{\delta A'}]\\ 
&\quad \cdot\ (e^{\mathfrak{D}_A}F_2[A'])
\end{split}
\end{equation}

\begin{equation}
e^{\mathfrak{D}_A}F_1[A]F_2[A]= \mathfrak{N} exp[-\frac{i}{2}\int \bar{Q}\cdot \hat{\bf{K}}^{-1}\cdot \bar{Q}+\frac{1}{2} tr \ln \hat{\bf{K}}]\cdot exp[\frac{i}{2}\int \frac{\delta}{\delta A}\cdot \hat{\bf{K}}^{-1}\cdot \frac{\delta}{\delta} - \int \bar{Q}\cdot \hat{\bf{K}}^{-1}\cdot \frac{\delta}{\delta A}]\cdot exp[L[A]]
\end{equation}

As one sees in above equation, all the explicit gauge dependent propagators cancels. This is gauge invariant by means of gauge-independence. It deserves to be emphasized that Gauge Independence is the strongest form of Gauge Invariance. Feynman had long hoped for this for QED.

\begin{figure}[h]
\centering
\includegraphics[width=4cm,clip]{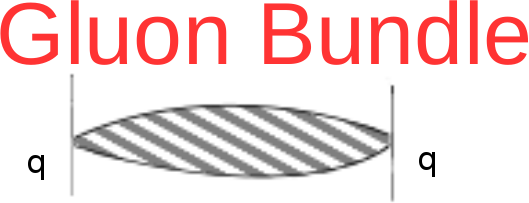}
\caption{A \it{Gluon Bundle}, GB, is the term $(f\cdot \chi)^{-1}$, representing the exchange of all gluons summed.}
\label{GB}       
\end{figure}

The $-\hat{\bf{K}}^{-1}$ above, also written as $(f\cdot\chi)^{-1}$ represents infinite gluon exchanges summed. This term is the \text{\it{Gluon Bundle}}, GB, exchanged between two quarks as shown in figure~\ref{GB}.

\begin{figure}[h]
\centering
\includegraphics[width=12cm,clip]{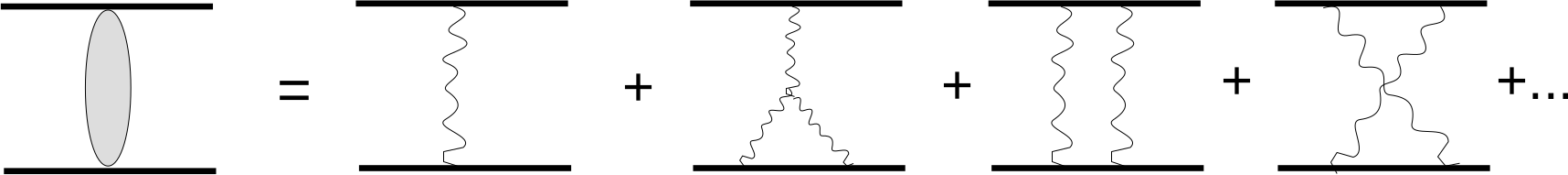}
\caption{A \it{Gluon Bundle}, GB, representing the exchange of all gluons summed.}
\label{GB2}       
\end{figure}

All the gaussian linkage operations can then be carried through exactly, corresponding to the summation of all gluons changed between any pair of quark (and/or anti-quark) lines, and including the cubic and quartic gluon interactions. See figure~\ref{GB2}. 

The result is explicit cancellation of all gauge-dependent gluon propagators, with resulting GF exhibiting Manifest Gauge Independence. One also finds a new, exact property of non-perturbative, gauge-invariant QCD, where the space-time coordinates of both ends of a GF are equal, modulo small uncertainties in their transverse coordinates. There is an Effective Locality between interacting quarks, and, changes all the remaining functional integrals from Schwinger into ordinary integrals. One can actually complete these integrals ~\cite{2}

The summation of all gluon exchanges, a Gluon Bundle, GB, is a Gaussian functional derivative on quarks.
figure~\ref{GB2}.

An important point is that quarks are never observed individually, and, thus, cannot have fixed coordinates. The correct coordinates for quarks include transverse quark fluctuations. We believe we know how to do this, the work is still underway for putting quark fluctuations from first principles. What we have done here is to introduce phenomenlogical transverse fluctuation amplitudes for every quark-gluon vertex, replacing the usual gluon-quark current interaction at the same space-time point $\int d^4 x \bar{\psi}(x)\gamma_{\mu}A_{\mu}^a(x)\tau_a \psi(x)$ by $\int d^2 x'_{\perp}\int d^4 x a(x_{\perp}-x'_{\perp})\bar{\psi}(x')\gamma_{\mu}\tau_aA_{\mu}^a(x)\psi(x')$, with $a(x_{\perp}-x'_{\perp})$ real and symmetric, and $x'_{\mu} = (x'_{\perp},x_L,x_0)$.

The probability of finding two quarks separated by a transverse (or impact parameter) distance is then $\varphi(b)=\int \frac{d^2q}{(2\pi)^2}e^{iqb}|\tilde{a}(q)|^2$.

We chose a deformed gaussian $\varphi(b)=\varphi(0)e^{(\mu b)^{2+\xi}}$  with deformation parameter $\xi$ real and small. A straight forward calculation yields, for small $\xi$, $V(r)\approx \xi \mu(\mu r)^{1+\xi}$. Perhaps it is important to emphasize again that all asymptotic quark states are hadronic bound states of quarks; and for such a bound state we can specify longitudinal and time coordinates, but not transverse coordinates since they are always fluctuating. The conventional "static quark" approximation used in model binding potential calculations in all non-perturbative amplitudes are plagued with divergences. All non-perturbative amplitudes are plagued with absurdities without taking such "transverse imprecision" into account. 

Substituting our potential into a Schrodinger binding equation, using the "quantic" approximation then yields $\mu \sim m_{\pi}$, with $\xi \approx 0.1$. This is sensible since the maximum fluctuations should be less than $m_{\pi}^{-1}$. 

Our results encompasses two different lattice calculations, $V \sim r$ and $V \sim r \ln (r)$. All lattice and other model calculations of $q-\bar{q}$ binding correspond to an amplitude containing only one or two Casimir $SU(3)$ invariants, $C_2$ or $C_3$, whereas our amplitude contains both~\cite{4}. We used the well-known half century old Eikonal function relation with potential. The minimum bound state energy for the pion shows that most of the pion's mass comes from the gluons forming the GB and relatively little from the quark masses.

\section{Nuclear physics from QCD}

\begin{figure}[h]
\centering
\includegraphics[width=7cm,clip]{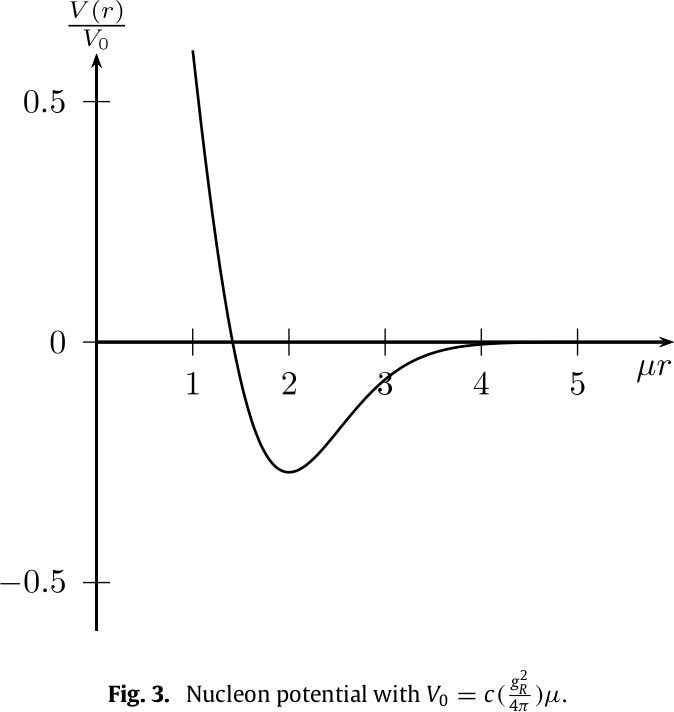}
\caption{Nucleon Nucleon binding is mediated by exchange of Gluon Bundles with quark loop that is able to stretch to distances greater than that of a pion.}
\label{potential}       
\end{figure}

In the same light, nucleon binding is examined. Here is the first (to our knowledge) example of nucleon binding, for a model deuteron, from basic QCD. We have performed a qualitative model, without electrical charge and nucleon spins which can always be added in, to describe the essence of Nuclear Physics. Assuming an average quark for ease of calculation, an attractive potential is obtained~\cite{5}.
Quark binding takes place for $r_{ij} \approx m_{\pi}^{-1}$, but for nucleon binding that takes place at larger distances, extraction and regularization of the logarithmic UV divergence loop will contribute two essential features. 1) The loop stretches, so distances larger than $m_{\pi}^{-1}$ can easily enter. 2) It provides a crucial change of sign for the effective n-n binding potential, figure~\ref{potential},~\cite{3}.
We expect and hope that nuclear physicists will employ such effective potentials to discuss heavy nuclei. 

All the basic radiative correction structure of non-perturbative QCD comes from interacting closed-quark-loops with GB's. A single dressed quark has an amplitude proportional to 

\begin{equation}
N \int d[\chi] e^{i\frac{\chi^2}{4}}(det(gf\cdot \chi)^{-\frac{1}{2}})e^{\hat{\mathfrak{D}_A}}\cdot G_c[A]e^{L[A]}|_{A\rightarrow 0},
\end{equation}

While two scattering quarks are described by 

\begin{equation}
N \int d[\chi] e^{i\frac{x\chi^2}{4}}(det(gf\cdot \chi)^{-\frac{1}{2}})e^{\hat{D}_A}G_c^{(1)}[A]G_c^{(2)}[A]e^{L[A]}|_{A\rightarrow 0},
\end{equation}

where $\hat{\mathfrak{D}}_A=\frac{i}{2}\int \frac{\partial}{\partial A}(gf\cdot \chi)^{-1}\frac{\partial}{\partial A}$.

\begin{figure}[h]
\centering
\includegraphics[width=12cm,clip]{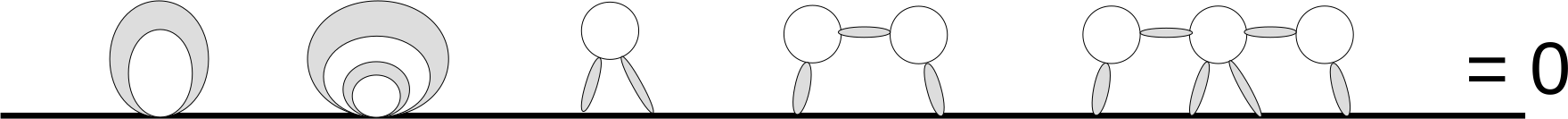}
\caption{All self energy graphs with Gluon Bundles are zero, 0.}
\label{quark_renormalization}       
\end{figure}

Every GB exchanged is represented by the linkage operator connecting the two $G_c[A]$'s to each other, and the $G_c[A]$'s to $L[A]$. Explicit calculation shows that all self-energy graphs vanish either by asymmetry of the $(f\cdot \chi)^{-1}$ color and indices or by explicit loop integration. See figure~\ref{quark_renormalization}
Non-perturbative QCD turns out to be far simpler than QED.

\begin{figure}[h]
\centering
\includegraphics[width=4cm,clip]{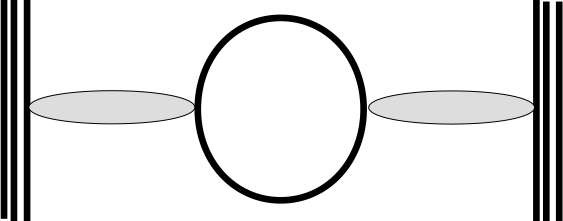}
\caption{Nucleon Nucleon binding is mediated by exchange of Gluon Bundles with quark loop that is able to stretch to distances greater than that of a pion.}
\label{nucleon}       
\end{figure}

\section{QCD renormalization}

The radiative corrections of QCD enter when there is momentum transfer between one quark and another quark, where momentum transfer passes through intemediate GBs and/or closed quark loops. 

\begin{figure}[h]
\centering
\includegraphics[width=7cm,clip]{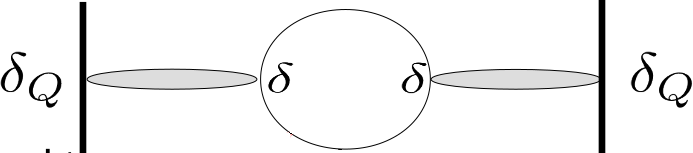}
\caption{We chose a renormalization scheme where two connections of Gluon Bundles, $\delta$, multplied by a quark loop with logarithmic divergence, $\ell$ is set to finite quantity $\kappa$ to be determined by experiments. }
\label{deltalkappa}       
\end{figure}

We use an exact functional cluster expansion described in Chapter 2.5 of~\cite{9}. In our particular choice of renormalization, we choose $\delta^2 \ell = \kappa$, where $\delta$ represent point where GB connects to a quark loop, $ell$ has the expected UV log divergence and $\kappa$ is a finite positive constant, figure~\ref{deltalkappa}  

\begin{figure}[h]
\centering
\includegraphics[width=12cm,clip]{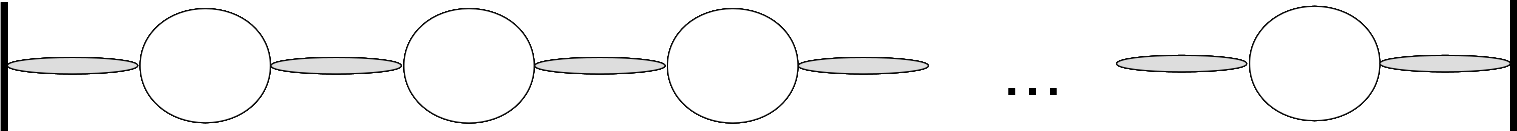}
\caption{We set $\delta^2 \cdot \ell = \kappa$, where $\kappa$ is finite and to be determined by experiments. This simplifies all posible loop connections with Gluon Bundles to only the straight chains. This is the first attempt and, thus far, compares well with experiments.}
\label{chainloop}       
\end{figure}

With this particular choice if renormalization, GB chain graphs are non-zero. All other closed loops entering into the functional cluster expansion vanish. These GB chain graphs form a geometric series which can be summed, and is everywhere finite. See figure~\ref{chainloop}.

This puts us in a position where we can compare with high energy experiment events. 

\section{Comparison with High Energy elastic pp-scattering experiments}

\begin{figure}[h]
\centering
\includegraphics[width=12cm,clip]{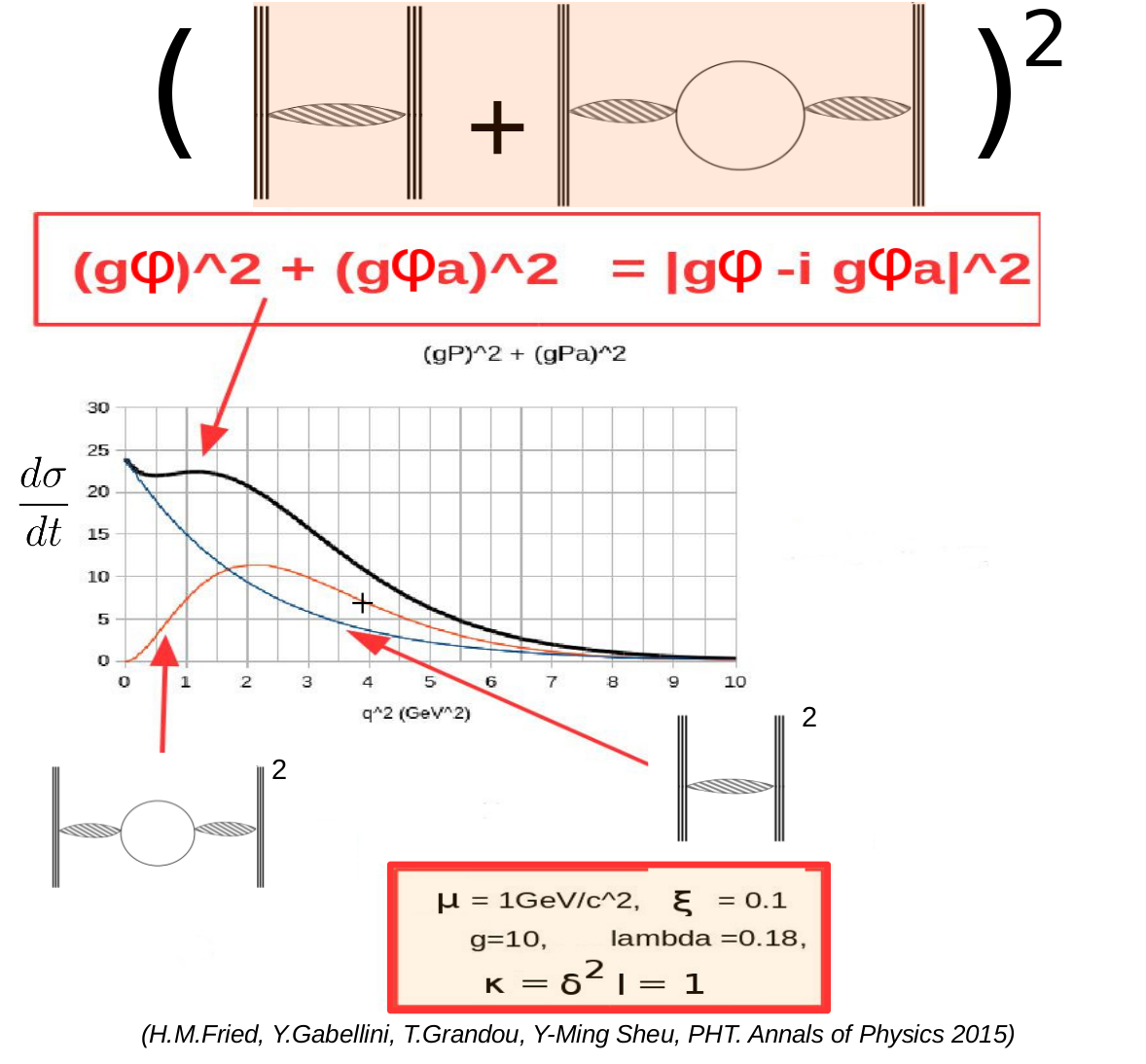}
\caption{Elastic differential cross-section is calculated from exchange of Gluon Bundles between hadrons and exchange of Gluon Bundles between hadrons with additional quark loop. We are using one loop as first calculation. Infinitely many loop chains is already calculated and will be put in in work to appear.}
\label{scatteringamplitude}       
\end{figure}

 In high energy hadron scattering, pp-scattering in particular, there has always been a 'diffraction dip' that was difficult to explain and certainly not from first principles. There are form factors, or other methods attempted. For us, we calculate the differential cross section from, intuitively trying a GB exchanged between two hadrons (3 quarks that are not breaking up) and a GB-virtual quark loop-GB configuration. Just to get an intuitive understanding. Low and behold, we have a diffraction dip as seen in experiments. It comes from quark-GB-quark and quark-GB-virtualquarkloop-GB-quark, figure~\ref{scatteringamplitude}, ~\cite{8}.

The contribution from purely GB's exchanged between two nucleons provide an amplitude with exponential fall off, while the one-loop-term provides a rising function, figure~\ref{scatteringamplitude}, that, when both combined explains the diffraction dip. The exact form that includes infinite sum of all chain loops are currently underway. We expect favorable comparison with the experimental measurements ~\cite{isr}.

\begin{figure}[h]
\centering
\includegraphics[width=10cm,clip]{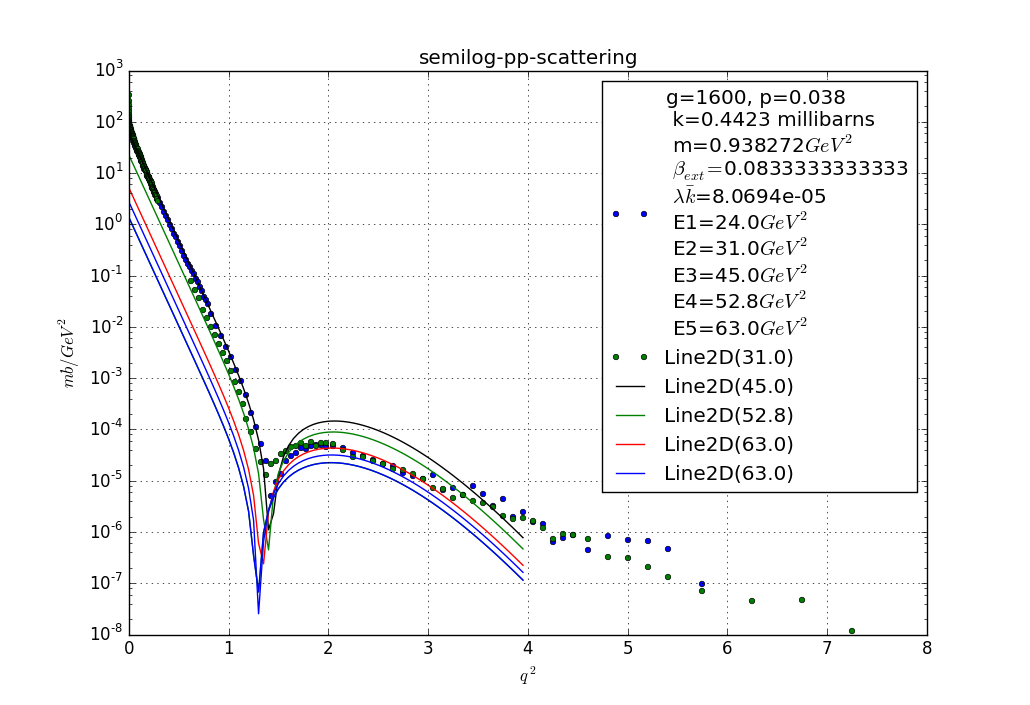}
\caption{Early comparisons of Gluon Bundle exchanges and the one-loop-term amplitude compares well with Intersecting Storage Ring data of elastic pp-scattering. Data points are in small circular points. Our calculations are solid lines. There is the expected movement of the dip to smaller $q^2$ as energy, $\sqrt{s}$ is increased. \cite{isr} }
\label{data}       
\end{figure}

The $\delta_q$ in our amplitude that connects to physical quark lines have units of time, giving us inverse eenergy relation. Raising $1/E$ to the first power provided too strong an energy dependence for positions of the dips. With $\delta_q$ proportional to $(1/m)(m/E)^{p}$, the power $p$ can be deduced from data. See figure~\ref{data}.

We have 3 parameters, coupling $g$, $\lambda\kappa$ and $p$ that will be determined from experiment. 
More detailed analysis is currently underway and, of course, to data at $7\ TeV$ and above.

\section{Acknowledgements}
This publication was made possible through the support of a Grant from the Julian Schwinger Foundation. The opinions expressed in this publication are those of the authors and do not necessarily reflect the views of the Julian Schwinger Foundation. We especially wish to thank Mario Gattobigio for his many kind and informative conversations relevant to the Nuclear Physics aspects of our work. It is also a pleasure to thank Mark Restollan, of the American University of Paris, for his kind assistance in arranging sites for our collaborative research when in Paris.

\end{document}